\documentclass[%
 aip,
 amsmath,amssymb,
 reprint,%
]{revtex4-1}
\usepackage{float}
\usepackage{graphicx}% Include figure files
\usepackage{dcolumn}% Align table columns on decimal point
\usepackage{bm}% bold math
\usepackage[mathlines]{lineno}% Enable numbering of text and display math
\usepackage[utf8]{inputenc}
\usepackage[T1]{fontenc}
\usepackage{mathptmx}
\usepackage{float}\usepackage[utf8]{inputenc}
\usepackage[T1]{fontenc}
\usepackage{textcomp}
\usepackage{gensymb}
\usepackage{multirow}
\usepackage{here}
\usepackage{graphicx}
\usepackage[dvipsnames]{xcolor} %feel free to delete this; it's to put Leora's comments in a different color
  
\newcommand{\un}[1]{\ensuremath{\,\mathrm{#1}}}
\newcommand{\degC}[0]{\ensuremath{\degree\mathrm{C}}}

\begin{document}

\preprint{AIP/123-QED}

\title[REVIEW OF SCIENTIFIC INSTRUMENTS]{Mirror Furnace for Synchrotron Dark Field X-ray Microscopy Experiments}

\author{C. Yildirim}
  \altaffiliation[Present address: ]{LETI, CEA, 17 avenue des Martyrs, 
38054 Grenoble, France}
  \affiliation{Experiments Division, European Synchrotron Radiation Facility, 71 Avenue des Martyrs, CS40220, 38043 Grenoble Cedex 9, France.}
 \affiliation{OCAS, Pres. J.F. Kennedylaan 3, BE-9060 Zelzate, Belgium.}

\author{H. Vitoux}%
  % \altaffiliation[Present address: ]{Gone! (not our problem)}
  \affiliation{Experiments Division, European Synchrotron Radiation Facility, 71 Avenue des Martyrs, CS40220, 38043 Grenoble Cedex 9, France.}
  
 \author{L. E. Dresselhaus-Cooper}%
   \affiliation{Lawrence Livermore National Laboratory, Physics and Life Sciences, Physics Division, 7000 East Avenue, L-487, Livermore, CA, USA, 94550.}  

\author{R. Steinmann}
  % \altaffiliation{Retired, but who cares?}
  \affiliation{Experiments Division, European Synchrotron Radiation Facility, 71 Avenue des Martyrs, CS40220, 38043 Grenoble Cedex 9, France.}
 
\author{Y. Watier} % probably did more work on this than Steinmann
   \affiliation{Experiments Division, European Synchrotron Radiation Facility, 71 Avenue des Martyrs, CS40220, 38043 Grenoble Cedex 9, France.}
   
\author{P. K. Cook}
  \altaffiliation[Present address: ]{Institut für Physik und Materialwissenschaft (IPM), University of Natural Resources and Life Sciences, Peter-Jordan-Stra{\ss}e 82, 1190 Vienna, Austria}
  \affiliation{Experiments Division, European Synchrotron Radiation Facility, 71 Avenue des Martyrs, CS40220, 38043 Grenoble Cedex 9, France.}

\author{M. Kutsal}
 \affiliation{Experiments Division, European Synchrotron Radiation Facility, 71 Avenue des Martyrs, CS40220, 38043 Grenoble Cedex 9, France.}%
 \affiliation{Department of Physics, Technical University of Denmark, 2800 Kgs.~Lyngby, Denmark.}
 
\author{C. Detlefs}
 \affiliation{Experiments Division, European Synchrotron Radiation Facility, 71 Avenue des Martyrs, CS40220, 38043 Grenoble Cedex 9, France.}%

\date{\today}% It is always \today, today,

\begin{abstract}

We present a multi-purpose mirror furnace designed for synchrotron X-ray experiments. The furnace is optimized specifically for dark-field X-ray microscopy (DFXM) of crystalline materials at the beamline ID06 of the ESRF. 
%DFXM is a novel non-destructive diffraction-based technique for 3D mapping of orientation and lattice strain within individual grains embedded in bulk samples with spatial and angular resolutions of 100\un{nm} and 0.0001$\degree$, respectively. 
The furnace can reach up to $\approx 1600\degC$ with stability better than $2\degC$, and heating and cooling rates up to $30\degree\mathrm{C/s}$. The contact-less design enables samples to be heated either in air or in a controlled atmosphere in a capillary tube. The temperature was calibrated via the thermal expansion of an $\alpha$-iron grain. Temperature profiles in the $y$ and $z$ axes were measured by scanning a thermocouple through the focal spot of the furnace.
In the current configuration of the beamline, the furnace can be used for DFXM, near-field X-ray topography, bright field X-ray nanotomography, high-resolution reciprocal space mapping, and limited powder diffraction experiments.
As a first application, we present a DFXM case study on isothermal heating of a commercially pure Al single crystal.

\end{abstract}

\maketitle

%\linenumbers

\section{\label{sec:int} Introduction}

Microstructure often governs the performance of engineering materials. Industrial applications use heat treatments to tune a material's microstructure for desired mechanical properties, as the microstructure of metals and alloys depend on the material's thermal history \cite{humphreys_2004}. Ceramics and optical materials also are often heat treated to modify the material performance, as seen in aerospace, energy, communication, and biomedical applications. A wide range of applications thus require \emph{in-situ} studies to understand the microscopic processes that occurs during heat treatment to design and develop engineering materials.

Many industrial materials have hierarchical structures of grains and domains that span several length scales, requiring that high-temperature studies of the material behavior consist of non-destructive measurements that can be accurate from 10\un{nm} to 1\un{mm}, with high angular resolution. Electron-based microscopy methods such as electron back-scatter diffraction (EBSD) and transmission electron microscopy (TEM) can probe several length-scales with a high spatial-resolution, however, the limited penetration depth of electron beams require very thin samples or invasive sectioning processes \cite{winther_critical_2004}. Synchrotron diffraction methods such as 3DXRD \citep{Schmidt229} and its near-field derivative diffraction contrast tomography (DCT) \citep{sun2018} can measure 3D information in a non-destructive manner, but cannot map highly deformed samples or nm-sized domains because they have overlapping peaks and insufficient spatial resolution , respectively. 

Dark-field X-ray microscopy (DFXM) is a new technique that non-destructively collects three-dimensional (3D) information about the material's strain and orientation at different length-scales \cite{Simons2015,Simons2016,Poulsen2017,Poulsen2018}. Analogous to dark-field electron microscopy, DFXM uses an objective lens to magnify features that diffract in a crystalline sample. Using high-energy synchrotron x-rays, the beam penetrates deep inside the crystal, resolving features that are deeply embedded inside the crystal. Full 3D mapping can be performed in minutes, with changes to the field of view and spatial resolution requiring only simple reconfiguration of the x-ray objective lens. Spatial resolution has reached 100\un{nm}, with an angular resolution of $0.001^\circ$. The high-resolution and high-sensitivity, coupled with the deep penetration-depth makes DFXM a unique tool to study the \textit{in-situ} behavior of crystalline engineering materials at high temperature.

Here, we present a non-contact optical furnace designed specifically to be compatible with DFXM experiments at the European Synchrotron Radiation Facility (ESRF).

\section{\label{sec:spec} Specifications}
\begin{figure*}[hbt!]
    \centering
    \includegraphics[width=2\columnwidth]{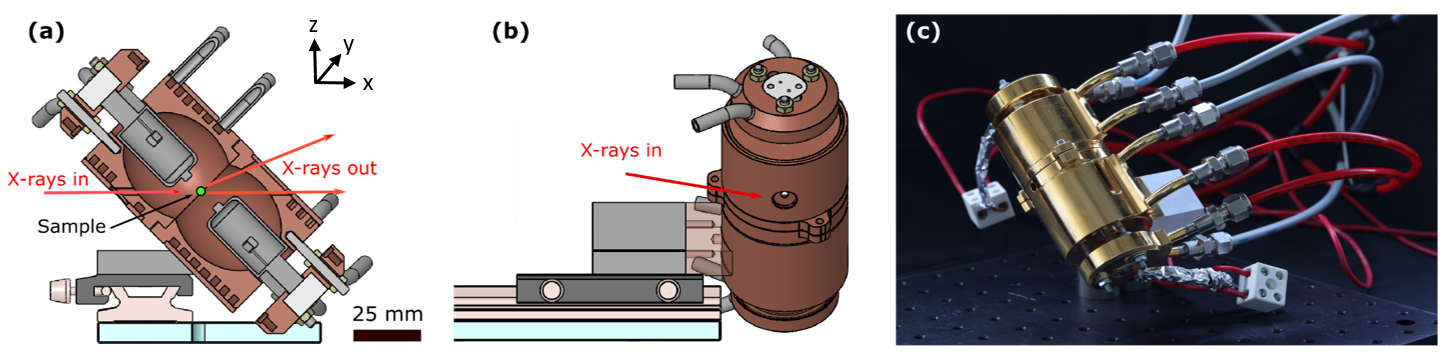}
    \caption{General layout of the furnace, (a) side view and (b) front view. The incident beam travels along the $x$-axis. (c) Photograph of the lamp furnace. The opening for the sample is on the left, and the slot for the scattered beam is on the right.}
    \label{fig:furnace_layout}
\end{figure*}

As the primary scientific goal of this furnace was originally to study heat treatments of metals, we have focused its calibrations on the temperature range relevant to process steel, namely, from room temperature to above $1200\degC$.
Within that range, we have explored fast rates of heating and cooling. The heating and cooling rates are important for industry applications that require isothermal studies. Cooling rates of 25-30 $\degree/s$ are typical for austenite to pearlite and/or martensite transition in many steel alloys \cite{bardelcik2010effect}. Therefore the furnace should achieve these cooling rates or more. 
\par
Thermal stability was important to investigate for both isothermal studies and to ensure that temperature
fluctuations do not alter the state of the sample during a DFXM scan. DFXM experiments use objective lenses with small effective apertures that are, therefore, highly sensitive to changes in the diffraction angle (from the $d$-spacing), which may be caused by thermal expansion or strain. 
\par

In designing this furnace for ID06, we also worked to ensure that the furnace could be compatible with the other X-ray techniques available there, including DFXM, bright-field tomography, near-field diffraction topography and high-resolution reciprocal space mapping. 
The furnace therefore has clear apertures to allow for the incident beam, diffracted beam, and/or transmitted beam of X-rays. The setup is designed to accommodate the current setup's diffraction along vertical the scattering plane \cite{Poulsen2018}, with scattering angles up to $2\theta \approx 30 \degree$. Finally, the furnace is designed to permit a full $360\degree$ sample rotation about the $y$ axis, which allows experiments to rock the crystal along one rotational axis for diffraction in the vertical geometry \cite{Poulsen2018}. 

As this work focuses on embedded structures within the material, the X-ray measurements are performed in transmission (Laue) geometry with photon energies in the range of 15--35\un{keV}. Typical samples have a size of 0.5--2\un{mm} in the $x$ and $z$ directions, and a length of at least several mm along the $y$ axis. The samples were mounted inside a quartz capillary that was filled either in air or in a protective atmosphere. 

The primary constraint in designing the furnace was the space available at the sample stage at ID06 \cite{Kutsal2019}, specifically, the distance between the mounting surface and the goniometer's center of rotation (i.e.~the sample position). The mounting surface of the goniometer is on the negative $y$ side (right side in Fig.~\ref{fig:furnace_layout}(b)).
Furthermore, the furnace should be small enough to fit a near-field camera  for sample alignment and near-field X-ray topography, with the camera sitting 50--100\un{mm} downstream from the sample.

\section{\label{sec:implement}Implementation}

To accommodate the specifications and constraints listed above, we chose a contact-less optical furnace, where the furnace and sample are mounted independently. This has the several benefits:

\begin{itemize} 
\item Heat transfer from the furnace to the goniometer is reduced, thus minimizing long-timescale thermal drift of the sample position.
\item The furnace and the associated power cables and cooling water lines do not increase the mechanical load on the goniometer stages.
\item Potential vibrations of the furnace (e.g.~due to water cooling) are not transmitted to the sample.
\item The furnace remains stationary during X-ray experiments, even when the sample is rotated or moved, minimizing the necessary size of the openings for the incident and diffracted beam. This reduces power loss, increasing the efficiency of the furnace and minimizing inhomogeneities in the heating.
\item As the furnace is mounted on a separate motorized $y$ translation stage, the furnace may be retracted without repositioning the sample. This can be helpful to access a wider range of scattering angles, as has been performed in interrupted annealing measurements with 3DXRD \cite{Mavrikakis2019}. 
\item The sample's temperature can be changed rapidly, as only the lamps and sample are "hot,"{i.e.~the thermal inertia of the system is very small. In principle, the rapid temperature response makes PID regulation easy. }

\end{itemize}

A schematic layout of the furnace is shown in Fig.~\ref{fig:furnace_layout}. Two lamps are used for the heating element in this design, as preliminary tests with a single bulb did not reach the desired temperatures. The bulbs are each positioned in a reflective cavity with a shape defined by two intersecting ellipsoids. The sample is positioned at the shared central focus, and the bulbs sit at the other two foci, as shown in Fig.~\ref{fig:furnace_layout}. The reflectors are machined from copper and they may be cooled by circulating either air or water in integrated cooling channels to regulate the temperature. After polishing the reflecting surfaces, the copper parts were gold plated.

The heating elements in the furnace are two identical 400-watt halogen light bulbs. The lamps are connected in parallel to a Delta Elektronika DC power supply, operated at constant power via a software PID loop. The required power for a given set-point temperature is obtained from a look-up table. As the sample's temperature can be varied with the beamline control software, data can be acquired automatically with slow, step-wise temperature ramps or interrupted annealing over a precise duration.
Different samples may require a different power to achieve the same set-point temperature due to deviations in it's surface reflectivity and absorption cross-section \cite{koohpayeh}. 

Samples can be inserted horizontally (perpendicular to the scattering plane) through an opening with $\approx 5\un{mm}$ diameter. The distance from the outside edge of the furnace to the sample position is $\approx 27\un{mm}$. The size of this opening allows the samples to be sealed in capillaries with a desired atmosphere, if necessary. The aperture for the incident X-rays has diameter of 2.5\un{mm}, matching the field of view of the near-field camera. On the downstream side, a vertical slot with 2.5\un{mm} width allows beams with vertical scattering angle up to 30$\degree$ to exit the furnace. The axis of the double ellipsoid furnace is inclined $45\degree$ towards the incident beam to maximize the accessible range of scattering angles, as shown in Fig.~\ref{fig:furnace_layout}.

The integrated cooling channels ensure that the outside temperature of the furnace remains below $20\degC$, even when it operates at temperatures above $1000\degC$. Parasitic heating of the surroundings is
thus limited is thus limited to radiation that escapes through the beam apertures, the sample aperture, and near the bases of the two lamps.
The cooling channels allow for safe operation of equipment surrounding the furnace, including the goniometer, sample stage, near-field camera, and others.

There are some disadvantages of the chosen geometry:
\begin{itemize}

\item The focal point of the furnace is located 27\un{mm} from the outer edges of the furnace. The sample mount must therefore be long and thin to position the material at the center of the furnace. As most of this support is heated, the sample position may require adjustments after large changes in temperature to correct for thermal expansion of the support(s).  

\item The small diameter of the sample aperture (5\un{mm}) strongly limits the range of sample rotations about the $x$ (incident beam) and $z$ (vertical) axes. Increasing the accessible angular range would require a larger lateral opening in the furnace, which would leak significantly more heat into the goniometer. These changes could also destabilize the sample position, causing asymmetry and inhomogeneity in the radiation and resulting heat load. 

\item The sample's temperature is not measured directly. The heat transferred to the sample and therefore the resulting sample temperature depend the absorption cross-section of the sample. At a given power, a strongly absorbing sample may be much hotter than a highly reflective sample. This means that the relationship between the power to the lamps and the temperature at the sample must be independently calibrated, for example, using either thermal expansion measurements. 

\item The small thermal mass and ability to rapidly change the sample temperature reduce the system's short-term temperature stability.

\end{itemize}

\section{\label{sec:results}Results and Discussions}

\subsection{Characterization by thermocouple}

\begin{figure}[ht]
    \centering
    \includegraphics[width=1\columnwidth]{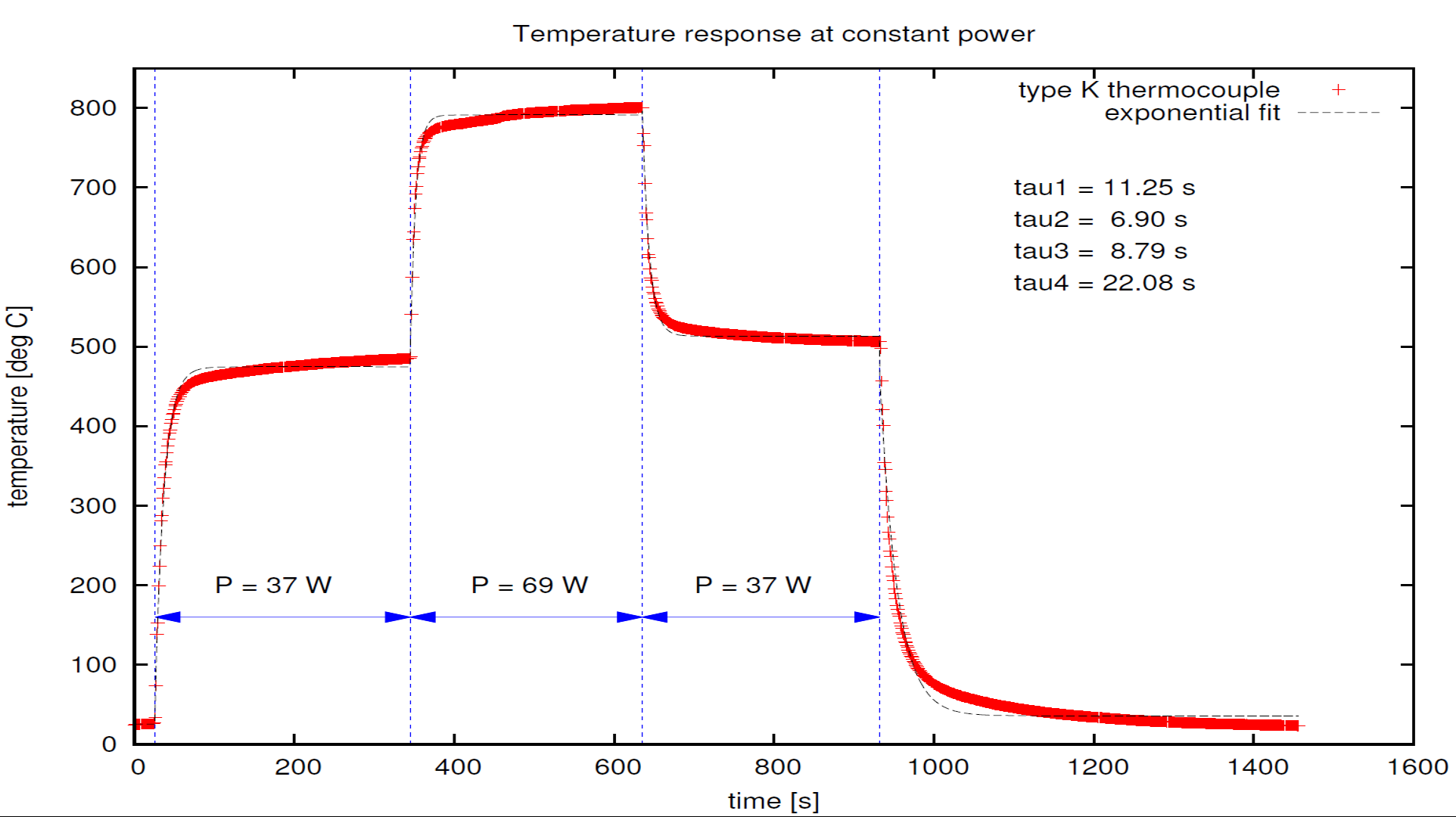}
    \caption{Temperature calibration using a K-type thermocouple showing heating and cooling rates for given set temperatures.}
    \label{fig:calibration}
\end{figure}

Figure \ref{fig:calibration} shows the temperature response of a K-type thermocouple located at the center of the furnace at different power values. 
5 minutes after reaching to the target temperature, the temperature stability of our furnace is better than $2 \degC$. Note that the measured temperatures are different at two plateaus of P = 37 W. This is due to the blackening of the thermocouple during the heating at P= 69 W, which resulted in further light absorption. therefore the measured temperature at P =37W for the second time is higher than that of the first time.  The heating and cooling times calculated using exponential fits to the temperature-time graph shown in Figure \ref{fig:calibration}.  Heating rates to a set temperature are faster than the cooling counter parts, due to the thermal mass of the furnace. Note in this experiment, we used air cooling. Nevertheless, the cooling rates can be increased by means of external air or gas blow using pipes mounted to the furnace and/or goniometer. 

%Our furnace allows heating and cooling rates up to $\approx 30 \degC$ /s.%

\begin{figure}
    \centering
    \includegraphics[width=0.5\textwidth]{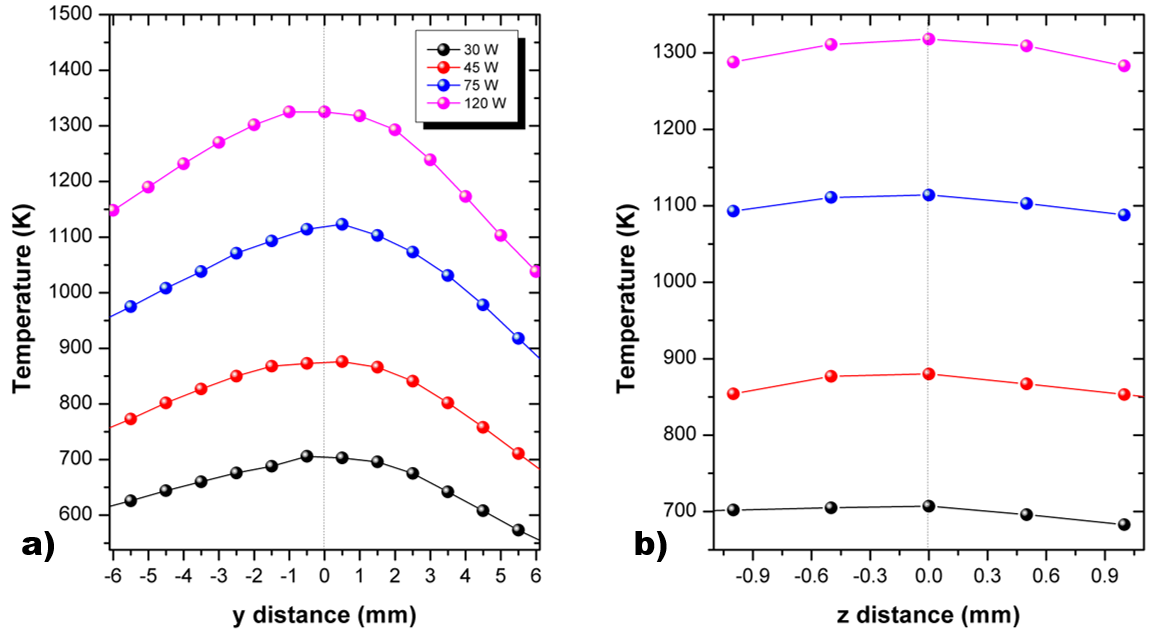}
    \caption{Spatial temperature distribution at different power settings in a) $y$ and b) $z$ directions as measured with a type K thermocouple. }
    \label{fig:position}
\end{figure}

Figure \ref{fig:position} shows the temperature distribution in $y$ and $z$ directions at different power of the lamps measured with a K-type thermocouple. The measured temperature increases as a function of power whereas it decreases as the thermocouple is further away from the center of the furnace. Both the temperature distribution is expected to be symmetrical for both $y$ and $z$ directions around the center.

\subsection{Characterization by lattice parameter}

We used a recrystallized $\alpha$ iron grain within a Fe-3$\%$Si sample with dimensions of 0.17 x 0.17 x 8 mm$^{3}$ inside a quartz capillary to calibrate the furnace. We probed the shift of the 110 Bragg peak in vertical direction with a photon energy of 17\un{keV} on a FReLoN CCD located 5090\un{mm} downstream of the sample. The \textit{d} spacing of 110 for each furnace power was used to calculate the temperature of the sample using the equation described elsewhere \cite{fe_calib}.

\subsection{Usage During a DFXM Experiment}

The furnace was tested during studies investigating the evolution of defects with temperature in single-crystals of commercially pure aluminum. The 0.5 x 0.5 x 20 mm$^{3}$ crystals were purchased from Surface Preparation Laboratories, where they were were cut from a larger sample and polished by chemical etching to reduce residual strains. Oriented along their crystallographic axes, the samples from a single slab of single-crystal commercially pure aluminum. No heat treatments were performed prior to slow-ramp heating of the samples in this experiment. The experiments were carried out on the dark-field x-ray microscope at the ESRF beamline ID06 \cite{Kutsal2019}. A Si(111) double crystal monochromator selected x-ray at photon energy 17 \un{keV} from the undulator source.

Before reaching the sample, the incident X-rays were passed through a 2D transfocator with compound refractive lenses (CRLs) totalling 8 Be lenslets (apex radius of curvature $R=200\un{\mu m}$), followed by a 1D condenser with 58 Be lenslets ($R=200\un{\mu m}$). A 300 $\times$ 2 $\mu$m (horizontal $\times$ vertical) spot of the sample was probed by the X-rays, which was then imaged with an X-ray objective comprised of 88 2D Be parabolic lenses ($R=50\un{\mu m})$ that was placed 274\un{mm} behind the sample, producing an effective focal length of 260\un{mm}. A far-field CCD camera was positioned 5364\un{mm} from the sample to capture the magnified image. The effective magnification was 18.5x, with a final effective pixel size of 75\un{nm/pixel} when combined with the scintillator-based optical components used with the detector. The incident X-ray line beam illuminated a $\approx  2\un{\mu m}$ high layer of the sample to capture section topographs. To map the relative axial strain, longitudinal ($\theta-2\theta$) scans were performed by collecting dark-field images across scans of the sample tilt, and scans of the objective and camera positions. We probed {002} Bragg reflection of the Aluminum crystal at $2\theta = \ 20.84 \degree$. Temperature was calibrated for the Al sample using the thermal expansion coefficients described in \cite{Al_calib}.

\begin{figure}[t]
    \centering
    \includegraphics[width=1\columnwidth]{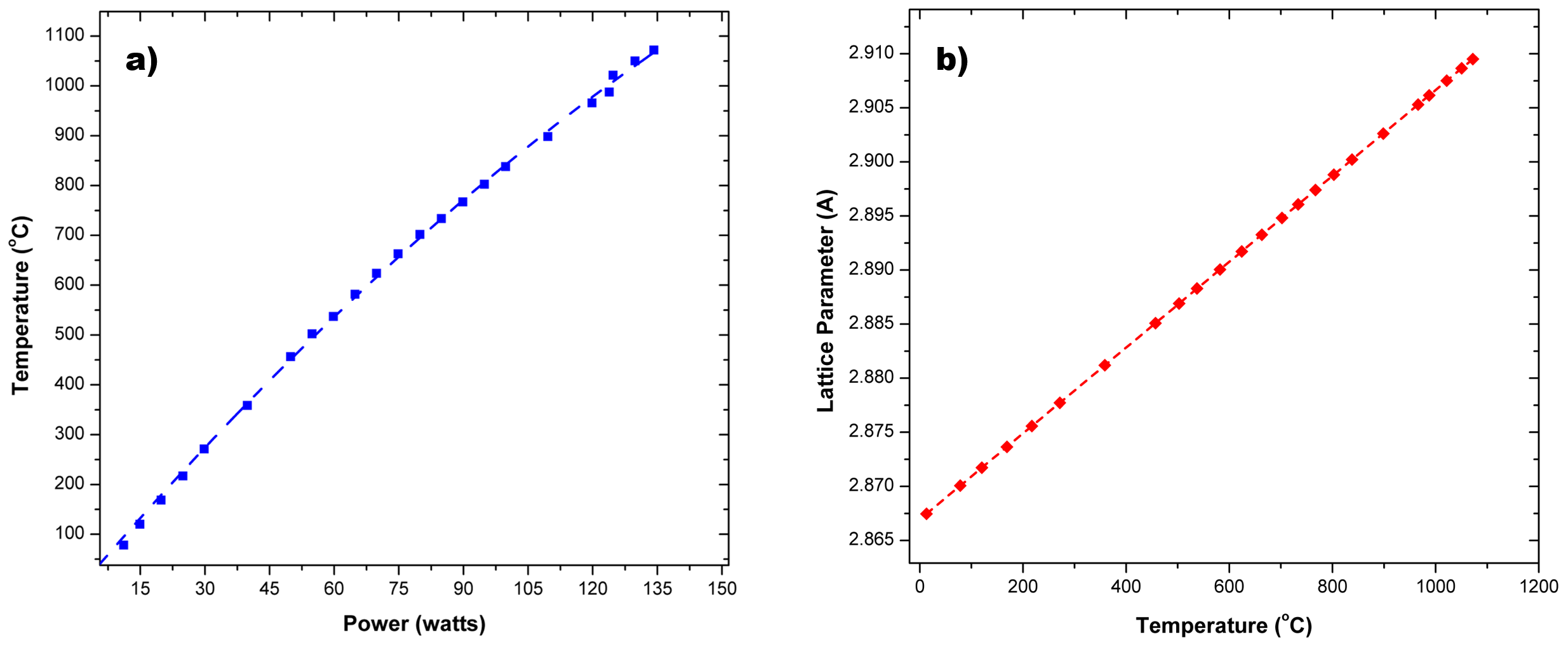}
    \caption{Temperature calibration curve, as measured based on thermal expansion of the lattice parameter in $\alpha$ iron. The 110 Bragg reflection of a grain was traced on a CCD located ~5 meters downstream of the sample. The plots show a) the sample temperature as a function of applied power, and b) the measured lattice parameter for each temperature upon heating. }
    \label{fig:calibration2}
    \vspace{-2mm}
\end{figure}
The images from this study reveal refinement of the microstructure with increasing temperature, resolving the temperature range required to anneal \cite{Williamson1966}. While we only observe poorly-resolved diffuse features at low T, as T increases, the features in each image clarify into linear structures. DFXM resolves local strain and misorientation in crystals that arise from long-range strain fields surrounding defects or grain boundaries. We interpret the low-temperature diffuse features to arise from a high density of defects, which are difficult to resolve from each other because their strain and misorientation overlap at our DFXM resolution. With this perspective, the change from diffuse to clearer features in our DFXM images suggest that we observe the decrease in dislocation density that is known to occur during annealing \cite{Mizunoa2005}. This indicates that the high-temperature features are likely single or small clusters of dislocations, as has been observed in diamond elsewhere \cite{Jakobsen2019}. The optical furnace in this work enabled the current experiment to probe a large region of sample that is evenly heated with limited thermal gradients. Similar studies can provide such insight to a host of other samples.

\begin{figure*}[!]
    \centering
    \includegraphics[width=2\columnwidth]{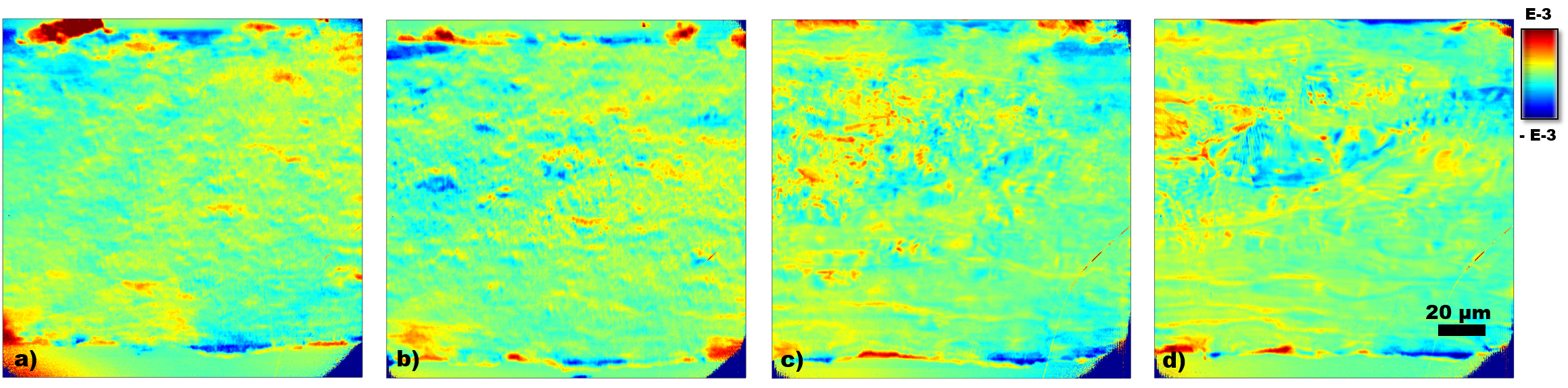}
    \caption{Reconstructed DFXM relative strain maps of Al single crystal during \textit{in-situ} isothermal heating at a) 155\degC{} b) 252\degC{} c) 416\degC{} d) 501\degC{}.  }
    \label{fig:strainmap}
\end{figure*}

\section{Conclusion}

We demonstrated a lamp furnace developed for DFXM studies at ID06 beamline at the ESRF. The furnace was calibrated using two distinct methods; by means of a K-type  thermocouple and tracing the lattice parameter change of an $\alpha$ iron grain. Temperatures up to 1600 C$\degree$ can be reached using the furnace, which makes it ideal to study steel alloys and ceramics. Even though the furnace was designed for DFXM studies, it is compatible with the other methods available at ID06 such as bright-field tomography, near-field diffraction topography and high-resolution reciprocal space mapping. We showed a case study of the DXFM of isothermal heating of pure Al single crystal. Axial strain maps showed structural modifications as a function of temperature. The current limitations of furnace can be given as the difficulty of the mosaicity scans due to the tilt range orthogonal to the rocking, and the impossibility of a full 3DXRD scan due to the limited size of the beam exit. Nevertheless, these studies are still possible for interrupted heating schemes by translating the furnace out without moving the sample.

\begin{acknowledgments}
We acknowledge the ESRF for provision of beamtime on beamline ID06. The DFXM demonstration of this work was performed under the auspices of the U.S. Department of Energy by Lawrence Livermore National Laboratory under Contract DE-AC52-07NA27344 and it was supported in part by LLNL, Lawrence Fellowship Program. C.~Y.~thanks Onderzoeks Centrum voor de Aanwending van Staal (OCAS), Belgium for the financial support of this project. 
\end{acknowledgments}

\section*{References}
\bibliography{golden_furnace}% Produces the bibliography via BibTeX.

\begin{thebibliography}{17}%
\makeatletter
\providecommand \@ifxundefined [1]{%
 \@ifx{#1\undefined}
}%
\providecommand \@ifnum [1]{%
 \ifnum #1\expandafter \@firstoftwo
 \else \expandafter \@secondoftwo
 \fi
}%
\providecommand \@ifx [1]{%
 \ifx #1\expandafter \@firstoftwo
 \else \expandafter \@secondoftwo
 \fi
}%
\providecommand \natexlab [1]{#1}%
\providecommand \enquote  [1]{``#1''}%
\providecommand \bibnamefont  [1]{#1}%
\providecommand \bibfnamefont [1]{#1}%
\providecommand \citenamefont [1]{#1}%
\providecommand \href@noop [0]{\@secondoftwo}%
\providecommand \href [0]{\begingroup \@sanitize@url \@href}%
\providecommand \@href[1]{\@@startlink{#1}\@@href}%
\providecommand \@@href[1]{\endgroup#1\@@endlink}%
\providecommand \@sanitize@url [0]{\catcode `\\12\catcode `\$12\catcode
  `\&12\catcode `\#12\catcode `\^12\catcode `\_12\catcode `\%12\relax}%
\providecommand \@@startlink[1]{}%
\providecommand \@@endlink[0]{}%
\providecommand \url  [0]{\begingroup\@sanitize@url \@url }%
\providecommand \@url [1]{\endgroup\@href {#1}{\urlprefix }}%
\providecommand \urlprefix  [0]{URL }%
\providecommand \Eprint [0]{\href }%
\providecommand \doibase [0]{http://dx.doi.org/}%
\providecommand \selectlanguage [0]{\@gobble}%
\providecommand \bibinfo  [0]{\@secondoftwo}%
\providecommand \bibfield  [0]{\@secondoftwo}%
\providecommand \translation [1]{[#1]}%
\providecommand \BibitemOpen [0]{}%
\providecommand \bibitemStop [0]{}%
\providecommand \bibitemNoStop [0]{.\EOS\space}%
\providecommand \EOS [0]{\spacefactor3000\relax}%
\providecommand \BibitemShut  [1]{\csname bibitem#1\endcsname}%
\let\auto@bib@innerbib\@empty
%</preamble>
\bibitem [{\citenamefont {Humphreys}\ and\ \citenamefont
  {Hatherly}(2004)}]{humphreys_2004}%
  \BibitemOpen
  \bibfield  {author} {\bibinfo {author} {\bibfnamefont {F.~J.}\ \bibnamefont
  {Humphreys}}\ and\ \bibinfo {author} {\bibfnamefont {M.}~\bibnamefont
  {Hatherly}},\ }\href {\doibase 10.1016/B978-0-08-044164-1.X5000-2} {\emph
  {\bibinfo {title} {Recrystallization {And} {Related} {Annealing}
  {Phenomena}}}},\ \bibinfo {edition} {2nd}\ ed.\ (\bibinfo  {publisher}
  {Elsevier},\ \bibinfo {year} {2004})\BibitemShut {NoStop}%
\bibitem [{\citenamefont {Winther}\ \emph {et~al.}(2004)\citenamefont
  {Winther}, \citenamefont {Huang}, \citenamefont {Godfrey},\ and\
  \citenamefont {Hansen}}]{winther_critical_2004}%
  \BibitemOpen
  \bibfield  {author} {\bibinfo {author} {\bibfnamefont {G.}~\bibnamefont
  {Winther}}, \bibinfo {author} {\bibfnamefont {X.}~\bibnamefont {Huang}},
  \bibinfo {author} {\bibfnamefont {A.}~\bibnamefont {Godfrey}}, \ and\
  \bibinfo {author} {\bibfnamefont {N.}~\bibnamefont {Hansen}},\ }\bibfield
  {title} {\enquote {\bibinfo {title} {Critical comparison of dislocation
  boundary alignment studied by {TEM} and {EBSD}: technical issues and
  theoretical consequences},}\ }\href {\doibase 10.1016/j.actamat.2004.05.050}
  {\bibfield  {journal} {\bibinfo  {journal} {Acta Materialia}\ }\textbf
  {\bibinfo {volume} {52}},\ \bibinfo {pages} {4437--4446} (\bibinfo {year}
  {2004})}\BibitemShut {NoStop}%
\bibitem [{\citenamefont {Schmidt}\ \emph {et~al.}(2004)\citenamefont
  {Schmidt}, \citenamefont {Nielsen}, \citenamefont {Gundlach}, \citenamefont
  {Margulies}, \citenamefont {Huang},\ and\ \citenamefont
  {Jensen}}]{Schmidt229}%
  \BibitemOpen
  \bibfield  {author} {\bibinfo {author} {\bibfnamefont {S.}~\bibnamefont
  {Schmidt}}, \bibinfo {author} {\bibfnamefont {S.~F.}\ \bibnamefont
  {Nielsen}}, \bibinfo {author} {\bibfnamefont {C.}~\bibnamefont {Gundlach}},
  \bibinfo {author} {\bibfnamefont {L.}~\bibnamefont {Margulies}}, \bibinfo
  {author} {\bibfnamefont {X.}~\bibnamefont {Huang}}, \ and\ \bibinfo {author}
  {\bibfnamefont {D.~J.}\ \bibnamefont {Jensen}},\ }\bibfield  {title}
  {\enquote {\bibinfo {title} {Watching the growth of bulk grains during
  recrystallization of deformed metals},}\ }\href {\doibase
  10.1126/science.1098627} {\bibfield  {journal} {\bibinfo  {journal}
  {Science}\ }\textbf {\bibinfo {volume} {305}},\ \bibinfo {pages} {229--232}
  (\bibinfo {year} {2004})},\ \Eprint
  {http://arxiv.org/abs/http://science.sciencemag.org/content/305/5681/229.full.pdf}
  {http://science.sciencemag.org/content/305/5681/229.full.pdf} \BibitemShut
  {NoStop}%
\bibitem [{\citenamefont {Sun}\ \emph {et~al.}(2018)\citenamefont {Sun},
  \citenamefont {Yu}, \citenamefont {Xu}, \citenamefont {Ludwig},\ and\
  \citenamefont {Zhang}}]{sun2018}%
  \BibitemOpen
  \bibfield  {author} {\bibinfo {author} {\bibfnamefont {J.}~\bibnamefont
  {Sun}}, \bibinfo {author} {\bibfnamefont {T.}~\bibnamefont {Yu}}, \bibinfo
  {author} {\bibfnamefont {C.}~\bibnamefont {Xu}}, \bibinfo {author}
  {\bibfnamefont {W.}~\bibnamefont {Ludwig}}, \ and\ \bibinfo {author}
  {\bibfnamefont {Y.}~\bibnamefont {Zhang}},\ }\bibfield  {title} {\enquote
  {\bibinfo {title} {3d characterization of partially recrystallized al using
  high resolution diffraction contrast tomography},}\ }\href@noop {} {\bibfield
   {journal} {\bibinfo  {journal} {Scripta Materialia}\ }\textbf {\bibinfo
  {volume} {157}},\ \bibinfo {pages} {72--75} (\bibinfo {year}
  {2018})}\BibitemShut {NoStop}%
\bibitem [{\citenamefont {Simons}\ \emph {et~al.}(2015)\citenamefont {Simons},
  \citenamefont {King}, \citenamefont {Ludwig}, \citenamefont {Detlefs},
  \citenamefont {Pantleon}, \citenamefont {Schmidt}, \citenamefont {St\"ohr},
  \citenamefont {Snigireva}, \citenamefont {Snigirev},\ and\ \citenamefont
  {Poulsen}}]{Simons2015}%
  \BibitemOpen
  \bibfield  {author} {\bibinfo {author} {\bibfnamefont {H.}~\bibnamefont
  {Simons}}, \bibinfo {author} {\bibfnamefont {A.}~\bibnamefont {King}},
  \bibinfo {author} {\bibfnamefont {W.}~\bibnamefont {Ludwig}}, \bibinfo
  {author} {\bibfnamefont {C.}~\bibnamefont {Detlefs}}, \bibinfo {author}
  {\bibfnamefont {W.}~\bibnamefont {Pantleon}}, \bibinfo {author}
  {\bibfnamefont {S.}~\bibnamefont {Schmidt}}, \bibinfo {author} {\bibfnamefont
  {F.}~\bibnamefont {St\"ohr}}, \bibinfo {author} {\bibfnamefont
  {I.}~\bibnamefont {Snigireva}}, \bibinfo {author} {\bibfnamefont
  {A.}~\bibnamefont {Snigirev}}, \ and\ \bibinfo {author} {\bibfnamefont
  {H.~F.}\ \bibnamefont {Poulsen}},\ }\bibfield  {title} {\enquote {\bibinfo
  {title} {Dark-field {X}-ray microscopy for multiscale structural
  characterization},}\ }\href {\doibase 10.1038/ncomms7098} {\bibfield
  {journal} {\bibinfo  {journal} {Nature Communications}\ }\textbf {\bibinfo
  {volume} {6}},\ \bibinfo {pages} {6098} (\bibinfo {year} {2015})}\BibitemShut
  {NoStop}%
\bibitem [{\citenamefont {Simons}\ \emph {et~al.}(2016)\citenamefont {Simons},
  \citenamefont {Jakobsen}, \citenamefont {Ahl}, \citenamefont {Detlefs},\ and\
  \citenamefont {Poulsen}}]{Simons2016}%
  \BibitemOpen
  \bibfield  {author} {\bibinfo {author} {\bibfnamefont {H.}~\bibnamefont
  {Simons}}, \bibinfo {author} {\bibfnamefont {A.~C.}\ \bibnamefont
  {Jakobsen}}, \bibinfo {author} {\bibfnamefont {S.~R.}\ \bibnamefont {Ahl}},
  \bibinfo {author} {\bibfnamefont {C.}~\bibnamefont {Detlefs}}, \ and\
  \bibinfo {author} {\bibfnamefont {H.~F.}\ \bibnamefont {Poulsen}},\
  }\bibfield  {title} {\enquote {\bibinfo {title} {Multiscale 3{D}
  characterization with dark-field x-ray microscopy},}\ }\href {\doibase
  10.1557/mrs.2016.114} {\bibfield  {journal} {\bibinfo  {journal} {MRS
  Bulletin}\ }\textbf {\bibinfo {volume} {41}},\ \bibinfo {pages} {454}
  (\bibinfo {year} {2016})}\BibitemShut {NoStop}%
\bibitem [{\citenamefont {Poulsen}\ \emph {et~al.}(2017)\citenamefont
  {Poulsen}, \citenamefont {Jakobsen}, \citenamefont {Simons}, \citenamefont
  {Ahl}, \citenamefont {Cook},\ and\ \citenamefont {Detlefs}}]{Poulsen2017}%
  \BibitemOpen
  \bibfield  {author} {\bibinfo {author} {\bibfnamefont {H.~F.}\ \bibnamefont
  {Poulsen}}, \bibinfo {author} {\bibfnamefont {A.~C.}\ \bibnamefont
  {Jakobsen}}, \bibinfo {author} {\bibfnamefont {H.}~\bibnamefont {Simons}},
  \bibinfo {author} {\bibfnamefont {S.~R.}\ \bibnamefont {Ahl}}, \bibinfo
  {author} {\bibfnamefont {P.~K.}\ \bibnamefont {Cook}}, \ and\ \bibinfo
  {author} {\bibfnamefont {C.}~\bibnamefont {Detlefs}},\ }\bibfield  {title}
  {\enquote {\bibinfo {title} {X-ray diffraction microscopy based on refractive
  optics},}\ }\href {\doibase https://doi.org/10.1107/S1600576717011037}
  {\bibfield  {journal} {\bibinfo  {journal} {Journal of Applied
  Crystallography}\ }\textbf {\bibinfo {volume} {50}},\ \bibinfo {pages}
  {1441--1456} (\bibinfo {year} {2017})}\BibitemShut {NoStop}%
\bibitem [{\citenamefont {Poulsen}\ \emph {et~al.}(2018)\citenamefont
  {Poulsen}, \citenamefont {Cook}, \citenamefont {Leemreize}, \citenamefont
  {Pedersen}, \citenamefont {Yildirim}, \citenamefont {Kutsal}, \citenamefont
  {Jakobsen}, \citenamefont {Trujillo}, \citenamefont {Ormstrup},\ and\
  \citenamefont {Detlefs}}]{Poulsen2018}%
  \BibitemOpen
  \bibfield  {author} {\bibinfo {author} {\bibfnamefont {H.~F.}\ \bibnamefont
  {Poulsen}}, \bibinfo {author} {\bibfnamefont {P.~K.}\ \bibnamefont {Cook}},
  \bibinfo {author} {\bibfnamefont {H.}~\bibnamefont {Leemreize}}, \bibinfo
  {author} {\bibfnamefont {A.~F.}\ \bibnamefont {Pedersen}}, \bibinfo {author}
  {\bibfnamefont {C.}~\bibnamefont {Yildirim}}, \bibinfo {author}
  {\bibfnamefont {M.}~\bibnamefont {Kutsal}}, \bibinfo {author} {\bibfnamefont
  {A.~C.}\ \bibnamefont {Jakobsen}}, \bibinfo {author} {\bibfnamefont {J.~X.}\
  \bibnamefont {Trujillo}}, \bibinfo {author} {\bibfnamefont {J.}~\bibnamefont
  {Ormstrup}}, \ and\ \bibinfo {author} {\bibfnamefont {C.}~\bibnamefont
  {Detlefs}},\ }\bibfield  {title} {\enquote {\bibinfo {title} {Reciprocal
  space mapping and strain scanning using x-ray diffraction microscopy},}\
  }\href {\doibase https://doi.org/10.1107/S1600576718011378} {\bibfield
  {journal} {\bibinfo  {journal} {Journal of Applied Crystallography}\ }\textbf
  {\bibinfo {volume} {51}},\ \bibinfo {pages} {1428--1436} (\bibinfo {year}
  {2018})}\BibitemShut {NoStop}%
\bibitem [{\citenamefont {Bardelcik}\ \emph {et~al.}(2010)\citenamefont
  {Bardelcik}, \citenamefont {Salisbury}, \citenamefont {Winkler},
  \citenamefont {Wells},\ and\ \citenamefont {Worswick}}]{bardelcik2010effect}%
  \BibitemOpen
  \bibfield  {author} {\bibinfo {author} {\bibfnamefont {A.}~\bibnamefont
  {Bardelcik}}, \bibinfo {author} {\bibfnamefont {C.~P.}\ \bibnamefont
  {Salisbury}}, \bibinfo {author} {\bibfnamefont {S.}~\bibnamefont {Winkler}},
  \bibinfo {author} {\bibfnamefont {M.~A.}\ \bibnamefont {Wells}}, \ and\
  \bibinfo {author} {\bibfnamefont {M.~J.}\ \bibnamefont {Worswick}},\
  }\bibfield  {title} {\enquote {\bibinfo {title} {Effect of cooling rate on
  the high strain rate properties of boron steel},}\ }\href@noop {} {\bibfield
  {journal} {\bibinfo  {journal} {International Journal of Impact Engineering}\
  }\textbf {\bibinfo {volume} {37}},\ \bibinfo {pages} {694--702} (\bibinfo
  {year} {2010})}\BibitemShut {NoStop}%
\bibitem [{\citenamefont {Kutsal}\ \emph {et~al.}(2019)\citenamefont {Kutsal},
  \citenamefont {Bernard}, \citenamefont {Berruyer}, \citenamefont {Cook},
  \citenamefont {Hino}, \citenamefont {Jakobsen}, \citenamefont {Ludwig},
  \citenamefont {Ormstrup}, \citenamefont {Roth}, \citenamefont {Simons},
  \citenamefont {Smets}, \citenamefont {Sierra}, \citenamefont {Wade},
  \citenamefont {Wattecamps}, \citenamefont {Yildirim}, \citenamefont
  {Poulsen},\ and\ \citenamefont {Detlefs}}]{Kutsal2019}%
  \BibitemOpen
  \bibfield  {author} {\bibinfo {author} {\bibfnamefont {M.}~\bibnamefont
  {Kutsal}}, \bibinfo {author} {\bibfnamefont {P.}~\bibnamefont {Bernard}},
  \bibinfo {author} {\bibfnamefont {G.}~\bibnamefont {Berruyer}}, \bibinfo
  {author} {\bibfnamefont {P.~K.}\ \bibnamefont {Cook}}, \bibinfo {author}
  {\bibfnamefont {R.}~\bibnamefont {Hino}}, \bibinfo {author} {\bibfnamefont
  {A.~C.}\ \bibnamefont {Jakobsen}}, \bibinfo {author} {\bibfnamefont
  {W.}~\bibnamefont {Ludwig}}, \bibinfo {author} {\bibfnamefont
  {J.}~\bibnamefont {Ormstrup}}, \bibinfo {author} {\bibfnamefont
  {T.}~\bibnamefont {Roth}}, \bibinfo {author} {\bibfnamefont {H.}~\bibnamefont
  {Simons}}, \bibinfo {author} {\bibfnamefont {K.}~\bibnamefont {Smets}},
  \bibinfo {author} {\bibfnamefont {J.~X.}\ \bibnamefont {Sierra}}, \bibinfo
  {author} {\bibfnamefont {J.}~\bibnamefont {Wade}}, \bibinfo {author}
  {\bibfnamefont {P.}~\bibnamefont {Wattecamps}}, \bibinfo {author}
  {\bibfnamefont {C.}~\bibnamefont {Yildirim}}, \bibinfo {author}
  {\bibfnamefont {H.~F.}\ \bibnamefont {Poulsen}}, \ and\ \bibinfo {author}
  {\bibfnamefont {C.}~\bibnamefont {Detlefs}},\ }\bibfield  {title} {\enquote
  {\bibinfo {title} {The {ESRF} dark-field x-ray microscope at {ID06}},}\ }in\
  \href@noop {} {\emph {\bibinfo {booktitle} {Proceedings of the 40th Ris{\o}
  International Symposium on "Metal Microstructures in 2D, 3D and 4D"}}}\
  (\bibinfo {year} {2019})\ \bibinfo {note} {to be published}\BibitemShut
  {NoStop}%
\bibitem [{\citenamefont {Mavrikakis}\ \emph {et~al.}(2019)\citenamefont
  {Mavrikakis}, \citenamefont {Detlefs}, \citenamefont {Cook}, \citenamefont
  {Kutsal}, \citenamefont {Campos}, \citenamefont {Gauvin}, \citenamefont
  {Calvillo}, \citenamefont {Saikaly}, \citenamefont {Hubert}, \citenamefont
  {Poulsen}, \citenamefont {Vaugeois}, \citenamefont {Zapolsky}, \citenamefont
  {Mangelinck}, \citenamefont {Dumont},\ and\ \citenamefont
  {Yildirim}}]{Mavrikakis2019}%
  \BibitemOpen
  \bibfield  {author} {\bibinfo {author} {\bibfnamefont {N.}~\bibnamefont
  {Mavrikakis}}, \bibinfo {author} {\bibfnamefont {C.}~\bibnamefont {Detlefs}},
  \bibinfo {author} {\bibfnamefont {P.}~\bibnamefont {Cook}}, \bibinfo {author}
  {\bibfnamefont {M.}~\bibnamefont {Kutsal}}, \bibinfo {author} {\bibfnamefont
  {A.}~\bibnamefont {Campos}}, \bibinfo {author} {\bibfnamefont
  {M.}~\bibnamefont {Gauvin}}, \bibinfo {author} {\bibfnamefont
  {P.}~\bibnamefont {Calvillo}}, \bibinfo {author} {\bibfnamefont
  {W.}~\bibnamefont {Saikaly}}, \bibinfo {author} {\bibfnamefont
  {R.}~\bibnamefont {Hubert}}, \bibinfo {author} {\bibfnamefont
  {H.}~\bibnamefont {Poulsen}}, \bibinfo {author} {\bibfnamefont
  {A.}~\bibnamefont {Vaugeois}}, \bibinfo {author} {\bibfnamefont
  {H.}~\bibnamefont {Zapolsky}}, \bibinfo {author} {\bibfnamefont
  {D.}~\bibnamefont {Mangelinck}}, \bibinfo {author} {\bibfnamefont
  {M.}~\bibnamefont {Dumont}}, \ and\ \bibinfo {author} {\bibfnamefont
  {C.}~\bibnamefont {Yildirim}},\ }\bibfield  {title} {\enquote {\bibinfo
  {title} {A multi-scale study of the interaction of sn solutes with
  dislocations during static recovery in $\alpha$-fe},}\ }\href {\doibase
  10.1016/j.actamat.2019.05.021} {\bibfield  {journal} {\bibinfo  {journal}
  {Acta Materialia}\ }\textbf {\bibinfo {volume} {174}},\ \bibinfo {pages}
  {92--104} (\bibinfo {year} {2019})}\BibitemShut {NoStop}%
\bibitem [{\citenamefont {Koohpayeh}\ \emph {et~al.}(2009)\citenamefont
  {Koohpayeh}, \citenamefont {Fort}, \citenamefont {Bradshaw},\ and\
  \citenamefont {Abell}}]{koohpayeh}%
  \BibitemOpen
  \bibfield  {author} {\bibinfo {author} {\bibfnamefont {S.}~\bibnamefont
  {Koohpayeh}}, \bibinfo {author} {\bibfnamefont {D.}~\bibnamefont {Fort}},
  \bibinfo {author} {\bibfnamefont {A.}~\bibnamefont {Bradshaw}}, \ and\
  \bibinfo {author} {\bibfnamefont {J.}~\bibnamefont {Abell}},\ }\bibfield
  {title} {\enquote {\bibinfo {title} {Thermal characterization of an optical
  floating zone furnace: A direct link with controllable growth parameters},}\
  }\href {\doibase https://doi.org/10.1016/j.jcrysgro.2009.02.017} {\bibfield
  {journal} {\bibinfo  {journal} {Journal of Crystal Growth}\ }\textbf
  {\bibinfo {volume} {311}},\ \bibinfo {pages} {2513 -- 2518} (\bibinfo {year}
  {2009})}\BibitemShut {NoStop}%
\bibitem [{\citenamefont {Basinski}, \citenamefont {Hume-Rothery},\ and\
  \citenamefont {Sutton}(1955)}]{fe_calib}%
  \BibitemOpen
  \bibfield  {author} {\bibinfo {author} {\bibfnamefont {Z.~S.}\ \bibnamefont
  {Basinski}}, \bibinfo {author} {\bibfnamefont {W.}~\bibnamefont
  {Hume-Rothery}}, \ and\ \bibinfo {author} {\bibfnamefont {A.}~\bibnamefont
  {Sutton}},\ }\bibfield  {title} {\enquote {\bibinfo {title} {The lattice
  expansion of iron},}\ }\href@noop {} {\bibfield  {journal} {\bibinfo
  {journal} {Proceedings of the Royal Society of London. Series A. Mathematical
  and Physical Sciences}\ }\textbf {\bibinfo {volume} {229}},\ \bibinfo {pages}
  {459--467} (\bibinfo {year} {1955})}\BibitemShut {NoStop}%
\bibitem [{\citenamefont {Wang}\ and\ \citenamefont {Reeber}(2000)}]{Al_calib}%
  \BibitemOpen
  \bibfield  {author} {\bibinfo {author} {\bibfnamefont {K.}~\bibnamefont
  {Wang}}\ and\ \bibinfo {author} {\bibfnamefont {R.~R.}\ \bibnamefont
  {Reeber}},\ }\bibfield  {title} {\enquote {\bibinfo {title} {The perfect
  crystal, thermal vacancies and the thermal expansion coefficient of
  aluminium},}\ }\href {\doibase 10.1080/01418610008212140} {\bibfield
  {journal} {\bibinfo  {journal} {Philosophical Magazine A}\ }\textbf {\bibinfo
  {volume} {80}},\ \bibinfo {pages} {1629--1643} (\bibinfo {year} {2000})},\
  \Eprint {http://arxiv.org/abs/https://doi.org/10.1080/01418610008212140}
  {https://doi.org/10.1080/01418610008212140} \BibitemShut {NoStop}%
\bibitem [{\citenamefont {Williamson}\ and\ \citenamefont
  {Smallman}(1966)}]{Williamson1966}%
  \BibitemOpen
  \bibfield  {author} {\bibinfo {author} {\bibfnamefont {G.~K.}\ \bibnamefont
  {Williamson}}\ and\ \bibinfo {author} {\bibfnamefont {R.~E.}\ \bibnamefont
  {Smallman}},\ }\bibfield  {title} {\enquote {\bibinfo {title} {Iii.
  dislocation densities in some annealed and cold-worked metals from
  measurements on the x-ray debye-scherrer spectrum},}\ }\href@noop {}
  {\bibfield  {journal} {\bibinfo  {journal} {Philosophical Magazine}\ }\textbf
  {\bibinfo {volume} {1}},\ \bibinfo {pages} {34--46} (\bibinfo {year}
  {1966})}\BibitemShut {NoStop}%
\bibitem [{\citenamefont {Mizunoa}\ \emph {et~al.}(2005)\citenamefont
  {Mizunoa}, \citenamefont {Yamamotoa}, \citenamefont {Morikawab},
  \citenamefont {Kugac}, \citenamefont {Okamotod},\ and\ \citenamefont
  {Hashimotoe}}]{Mizunoa2005}%
  \BibitemOpen
  \bibfield  {author} {\bibinfo {author} {\bibfnamefont {K.}~\bibnamefont
  {Mizunoa}}, \bibinfo {author} {\bibfnamefont {S.}~\bibnamefont {Yamamotoa}},
  \bibinfo {author} {\bibfnamefont {K.}~\bibnamefont {Morikawab}}, \bibinfo
  {author} {\bibfnamefont {M.}~\bibnamefont {Kugac}}, \bibinfo {author}
  {\bibfnamefont {H.}~\bibnamefont {Okamotod}}, \ and\ \bibinfo {author}
  {\bibfnamefont {E.}~\bibnamefont {Hashimotoe}},\ }\bibfield  {title}
  {\enquote {\bibinfo {title} {Vacancy generation mechanism at high
  temperatures in ultrahigh-purity aluminum single crystals with a low
  dislocation density},}\ }\href@noop {} {\bibfield  {journal} {\bibinfo
  {journal} {Journal of Crystal Growth}\ }\textbf {\bibinfo {volume} {275}},\
  \bibinfo {pages} {e1697--e1702} (\bibinfo {year} {2005})}\BibitemShut
  {NoStop}%
\bibitem [{\citenamefont {Jakobsen}\ \emph {et~al.}(2019)\citenamefont
  {Jakobsen}, \citenamefont {Simons}, \citenamefont {Ludwig}, \citenamefont
  {Leemreize}, \citenamefont {Porz}, \citenamefont {Detlefs},\ and\
  \citenamefont {Poulsen}}]{Jakobsen2019}%
  \BibitemOpen
  \bibfield  {author} {\bibinfo {author} {\bibfnamefont {A.~C.}\ \bibnamefont
  {Jakobsen}}, \bibinfo {author} {\bibfnamefont {H.}~\bibnamefont {Simons}},
  \bibinfo {author} {\bibfnamefont {W.}~\bibnamefont {Ludwig}}, \bibinfo
  {author} {\bibfnamefont {H.}~\bibnamefont {Leemreize}}, \bibinfo {author}
  {\bibfnamefont {L.}~\bibnamefont {Porz}}, \bibinfo {author} {\bibfnamefont
  {C.}~\bibnamefont {Detlefs}}, \ and\ \bibinfo {author} {\bibfnamefont
  {H.~F.}\ \bibnamefont {Poulsen}},\ }\bibfield  {title} {\enquote {\bibinfo
  {title} {Mapping of individual dislocations with dark field x-ray
  microscopy},}\ }\href {\doibase 10.1107/S1600576718017302} {\bibfield
  {journal} {\bibinfo  {journal} {J. Appl. Cryst.}\ }\textbf {\bibinfo {volume}
  {52}},\ \bibinfo {pages} {122} (\bibinfo {year} {2019})}\BibitemShut
  {NoStop}%
\end{thebibliography}%

\end{document}